\documentstyle[12pt]{article}
\begin{document}
\begin{center}
{\Large\bf  Zero-temperature dynamic transition
in the random field Ising model: A Monte Carlo study}
\end{center}
\vspace {0.4 cm}
\begin{center} {\large Muktish Acharyya$^*$} \end{center}
\vspace {0.3 cm}
\begin{center} {\it Institute for Theoretical Physics} \end{center}
\begin{center} {\it University of Cologne, 50923 Cologne, Germany} \end{center}
\vspace {1.0 cm}
\noindent {\bf Abstract:} The dynamics of a random (quenched) field
Ising model (in two dimension) at {\it zero temperature}
in the presence of an additional sinusoidally oscillating homogeneous
(in space) magnetic
field has been studied 
by Monte Carlo simulation using the Metropolis single spin flip
dynamics. 
The instantaneous magnetisation is found to be periodic with the same 
periodicity of the oscillating magnetic field. For very low values of amplitude
of oscillating field and the width of randomly quenched magnetic field, the
magnetisation oscillates asymmetrically about a nonzero value and 
the oscillation becomes
symmetric about a zero value for higher values of amplitude of oscillating
field and the width of the quenched disorder.
The time averaged magnetisation over a full
cycle of the oscillating magnetic field defines the dynamic order parameter.
This dynamic order parameter is nonzero for very low values of amplitude of
oscillating magnetic field and the width  of randomly quenched field. A phase
boundary line is drawn in the plane formed by the amplitude of the oscillating
magnetic field and the width of the randomly quenched magnetic field. A
tricritical point has been located, on the phase boundary line, which 
separates the nature (discontinuous/continuous) of the dynamic transition.

\vspace {1 cm}
\leftline {\bf Keywords: Random field Ising model,
Monte Carlo simulation}
\leftline {\bf PACS Numbers: 05.50 +q}
\vspace {2 cm}
\leftline {----------------------------}
\leftline {\bf $^*$E-mail:muktish@thp.uni-koeln.de}
\newpage

\leftline {\bf I. Introduction}

The Ising model in a quenched random field or the so called random field Ising 
model (RFIM) is an active field of modern research over a few 
decades. The behaviour of this model at zero temperature and the
zero temperature ferro-para phase transition has been studied 
widely \cite{rev}. Recently, a simple model was proposed \cite{sethna} for
hysteresis in magnetic model systems incorporating the return point memory
and Barkhausen noise \cite{brien}. For the smaller values of the quenched
disorder, the steady-state magnetisation has a jump discontinuity at a critical
value of the uniform external magnetic field and it vanishes
continuously for larger values of disorder strength. The transition from `jump'
to `no jump' in magnetisation was observed 
\cite{perko} at a critical value of the width
of the Gaussian disorder and the scaling behaviour was observed in the vicinity
of this critical value of the disorder. The exact solution \cite{shukla} of
this model in one dimension shows no jump discontinuity for Gaussian disorder.

Recently, Dhar et al \cite{dhar} studied the zero-temperature hysteresis in
the RFIM on a Bethe lattice. Surprisingly, 
they observed for coordination number $z = 3$ that
there is no jump discontinuity for any nonzero strength of Gaussian disorder.
However, for larger values of coordination number
and for very weak disorder, the magnetisation shows a jump discontinuity as a
function of external uniform field, which disappears for strong disorder.

All these transitions discussed above are static transitions characterised by
the steady state magnetisation. No attempt has been made to investigate the
dynamic transition in this model. The dynamic transition in the absence of
quenched random disorder and at finite (nonzero) temperatures are studied
widely [7-9]. 
For an introduction,
it would be convenient to review briefly the previous studies
on dynamic transition in the following paragraph.

The dynamic transition, in the kinetic Ising model in presence of a sinusoidally
oscillating magnetic field, was first studied by Tome
\& Oliveira \cite{tom}. They solved the mean-field dynamic equation for the
averaged magnetisation and defined the time averaged magnetisation over a full
cycle of the oscillating field as the dynamic order parameter. In the plane
formed by the temperature and the amplitude of the oscillating field, they
have drawn a dynamic phase boundary below which the dynamic phase is ordered
(nonzero value of the dynamic order parameter) and above which dynamic order
parameter vanishes characterising a disordered phase. They have also located
a tricritical point (TCP) on the phase boundary line which separates the nature 
(discontinuous/continuous) of this transition. However, this transition is not
truely dynamic in nature since it exists even in the static (zero frequency)
limit due to the absence of any fluctuation. The true dynamic transition (in
presence of thermal fluctuation) was studied extensively [8,9]. There are
some indications in the recent studies \cite{ma} that this transition posseses
some features analogous to 
equilibrium thermodynamic phase transitions. It may be
mentioned here that the evidence of dynamic transition, 
associated to the dynamic symmetry breaking, is found in the kinetic
Ising model in presence of a randomly varying (in time but uniform in space)
magnetic field \cite{ma1}.

In this paper, the dynamic transition has been studied, in the random field
Ising model driven by an (additional) sinusoidally varying (in time) 
homogeneous 
(in space) magnetic field, by Monte Carlo 
simulation using the Metropolis single spin-flip
dynamics at zero temperature. 
This has been organised as follows: in section II the model
and the simulation scheme are discussed, the simulation results are given
in section III and the paper ends with a summary of the work in section IV.

\bigskip
\leftline {\bf II. The model and simulation}

A square lattice of linear size $L$ is taken. Each site is labelled by
an integer $i$ and carries an Ising spin $S_i$ ($S_i = \pm1$) which interacts
with all its nearest neighbours (spins) with a ferromagnetic interaction
strength $J$. At each site $i$, there is a local {\it quenched} random field
$h_i$. The random fields 
${h_i}$ are assumed to be independent and identically distributed
random variables with a rectangular probability distribution $P(h_i)$. The
random field
$h_i$ can take any value from $-w/2$ to $+w/2$ with the same probability. The
width of the distribution is $w$.
In addition, there is a uniform (in space) magnetic field $h(t)$ which is
varying sinusoidally ($h(t) = h_0 \cos(\omega t)$) in time. The amplitude and
the frequency are denoted by $h_0$ and $\omega$ respectively. This kind of
model is described by the Hamiltonian
\begin{equation}
H = -J \sum_{<ij>} S_i S_j - \sum_i h_i S_i - h(t) \sum_i S_i,
\end{equation}
\noindent under the periodic boundary condition. For simplicity,
the interaction strength $J$ has been set equal to unity throughout the
study. 

The {\it zero-temperature} single spin-flip dynamics is specified by the
transition rates ($W$) \cite{kawasaki}
\begin{equation}
W(S_i \to -S_i) = \Gamma, ~~{\rm if~~} \Delta E \leq 0 {~~~\rm and~~}
W(S_i \to -S_i) = 0, {~~\rm otherwise}
\end{equation}
\noindent where $\Delta E$ is the change in energy due to spin flip. In 
words the algorithm is: never flip the chosen spin  
if this process would increase the energy and flip otherwise. We have
started with all spins are up ($S_i = +1$) as an initial condition and updated
the lattice sequentially using the above flipping algorithm. One such full
scan over the entire lattice consists a Monte Carlo 
step per spin (or MCS). The
instantaneous magnetisation (per site) $m(t)$ is easily calculated,
\begin{equation}
m(t) = {1 \over {L^2}} \sum_i S_i.
\end{equation}
\noindent After an initial transient period the intantaneous magnetisation
$m(t)$ has been found to be stabilised and 
periodic with the same periodicity of the applied
oscillating field. The dynamic order parameter $Q$, is defined as
\begin{equation}
Q = {{\omega} \over {2\pi}} \oint m(t) dt,
\end{equation}
\noindent which can be viewed as
the time averaged magnetisation over a full cycle of the
oscillating field. For a particular values of $h_0$, $\omega$ and $w$ the 
dynamic order parameter $Q$ is calculated by averaging over 20 different
random disorder (quenched) realisations.

\bigskip
\leftline {\bf III. Results}

The simulations are performed on a square lattice of linear size
$L = 100$ and a particular value of the frequency 
($\omega = 0.01 \times 2\pi$) of the oscillating
magnetic field. The time is measured in units of Monte Carlo steps per
spin or MCS and the values of random field and the oscillating field are
measured in units of interaction strength $J$.

It has been observed numerically that,
for fixed values of $h_0$ and $w$, 
the magnetisation becomes periodic (in time) with the same periodicity as
the applied sinusoidal magnetic field. For the smaller values of the quenched
disorder ($w = 8.0$) and the field amplitude ($h_0 = 0.5$), the magnetisation
oscillates asymmetrically about the zero line (Fig. 1a)
i.e., the system remains in a dynamically symmetry broken
phase. Consequently, the
hysteresis ($m-h$) loop 
resides on the upper half plane formed by $h(t)$ and
$m(t)$ (Fig. 1b). So, the time averaged magnetisation over a full cycle of
the oscillating field, the dynamic order parameter, is nonzero
in the symmetry broken phase. By increasing 
the field amplitude (for $h_0 = 2.0$) 
keeping $w$ fixed ($w = 8.0$), it was observed that the system 
acquires a 
dynamically symmetric phase, i.e., the magnetisation oscillates 
symmetrically about the zero line
(Fig. 1c). The hysteresis loop is also symmetric (Fig. 1d). Consequently, the
value of the dynamic order paramater $Q$ is zero in this dynamically
disordered (symmetric) phase.

In the dynamically disordered phase, the dynamic order parameter
$Q$ can be 
kept at zero in two ways, either by increasing the random field width $w$
for a fixed field amplitude $h_0$ or vice versa. 
So, in the plane formed by
the field amplitude ($h_0$) and the width ($w$) of the 
quenched disorder (random field), one can think of a
boundary line, below which $Q$ is nonzero and above
which it vanishes. Figure 2a displays such a phase boundary in the $h_0-w$
plane obtained by Monte Carlo simulation. 
The nature (discontinuous/continuous) of the transition depends on the
value of $w$ and $h_0$ on the phase boundary line. The transition across
the upper part of phase boundary line is discontinuous and it is continuous
for the rest part of the boundary. A tricritical point 
 on the phase boundary
line separates these natures. Figure 2b demonstrates two typical transitions
for two sets of values of $w$ and $h_0$ lying 
just in the left and right sides of the tricritical
point (TCP). In the case of discontinuous transition, the dynamic order
parameter $Q$ jumps to a small nonzero value and then vanishes continuously.
The uncertainty in the location of the TCP on the
phase boundary are shown by the circle enclosing it.

\bigskip
\leftline {\bf IV. Summary}

The {\it zero-temperature} dynamics of random field Ising model in
presence of an additional sinusoidally varying homogeneous field has been
studied by Monte Carlo simulation using the Metropolis single spin-flip
algorithm. The instantaneous magnetisation has been calculated and found
to be periodic with the same periodicity of the applied oscillating field.
For quite small values of amplitude of the oscillating field and the
width of randomly quenched field, the magnetisation oscillates 
asymmetrically about a nonzero value giving an asymmetric hysteresis loop,
and the systems remains in a dynamically ordered (nonzero dynamic order
parameter) symmetry broken phase. On the other hand, for quite large values
of disorder-strength and field-amplitude, the magnetisation oscillates
symmetrically about zero. As a consequence, the hysteresis loop is
symmetric and the system remains in a symmetric 
dynamically disordered (dynamic order
parameter vanishes) phase. The phase boundary line is sketched
in the plane formed by the disorder-strength and field-amplitude. The nature
(discontinuous/continuous) of the transition across the phase boundary
depends on the value of disorder strength. For low values of the disorder
strength the transition is discontinuous and becomes continuous for higher
values. A tricritical point  
(located on the phase boundary) separates the
nature of the transition.

Experimental evidence of a dynamic transition via a dynamical symmetry
(of the hysteresis loop) breaking has been found \cite{jiang} in the 
ultrathin ferromagnetic Co/Cu(001) sample by magneto-optic Kerr effect.
However, the phase boundary and its details (TCP) are not yet investigated
experimentally.

Recently, a theory of the hysteresis loop in ferromagnets controlled
by the domain wall motion has been developed \cite{nater}, where the domain
walls are considered as plane or linear interfaces moving in a random medium
under the action of the external {\bf ac} magnetic field.

\bigskip

\noindent {\bf Acknowledgments:} This work is supported by {\bf SFB 341}.

\newpage

\begin{center} {\bf Figure Captions} \end{center}

\bigskip

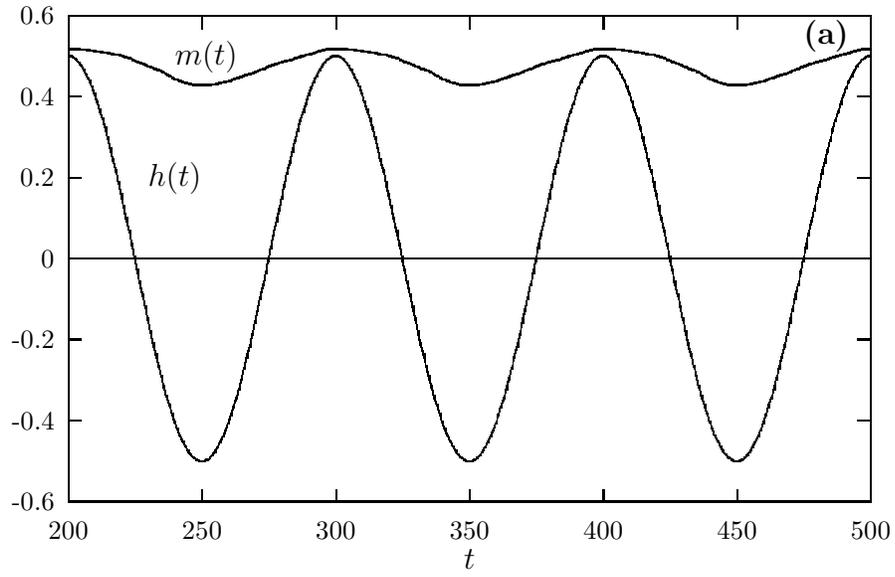
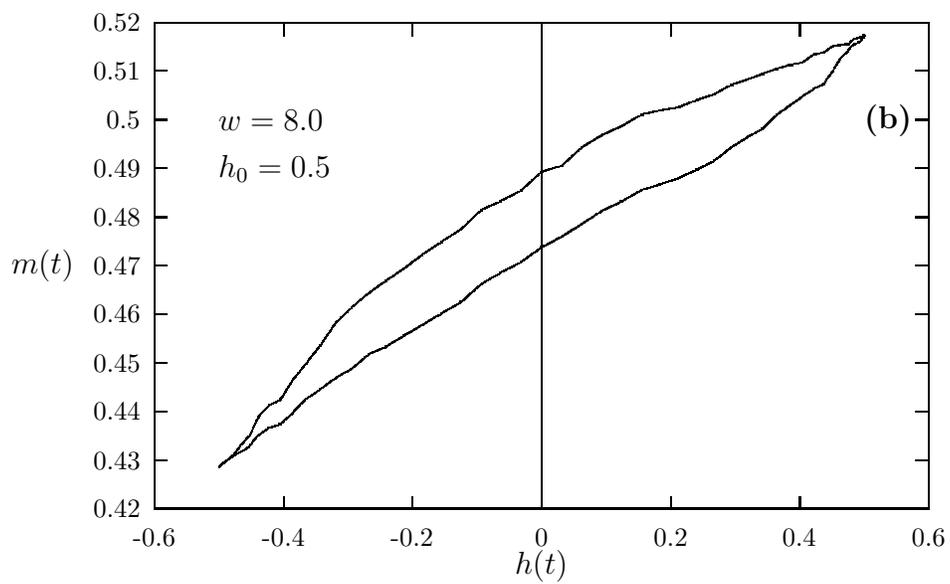
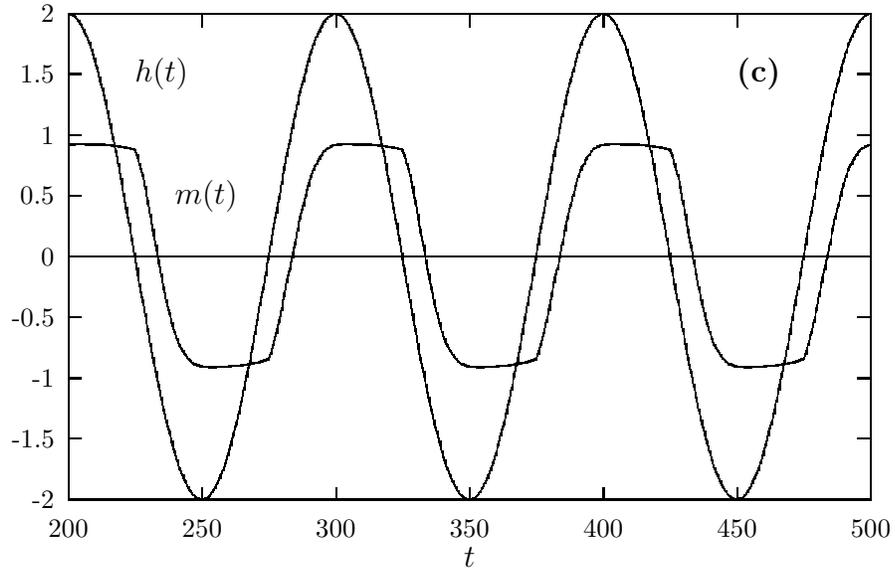
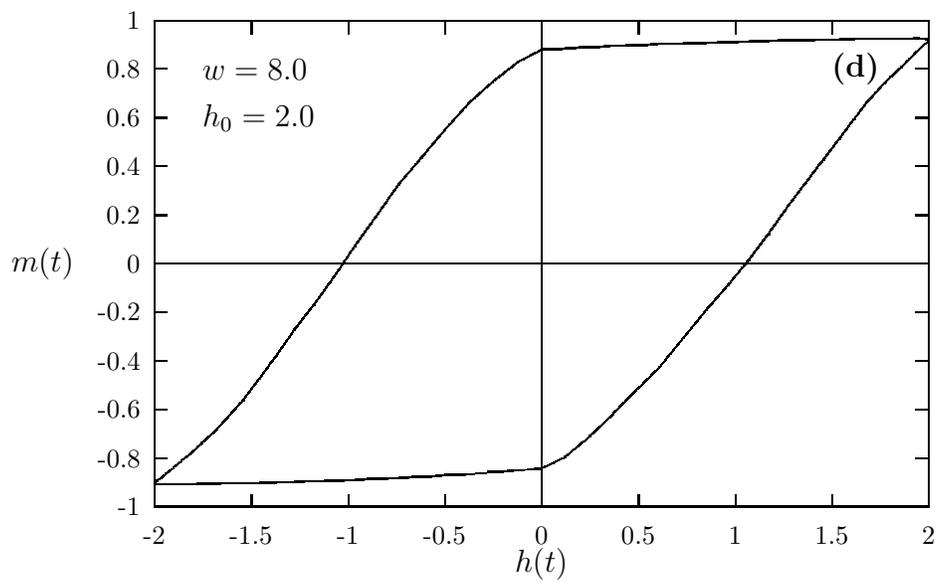
\noindent {\bf Fig. 1.} (a) The time variation of magnetisation $m(t)$ and
the field $h(t)$ for $h_0 = 0.5$ and $w = 8.0$, (b) the corresponding
hysteresis ($m(t)-h(t)$) loop (see the scale of y-axis), (c) the time
variations of magnetisation $m(t)$ and $h(t)$ for $h_0 = 2.0$ and $w = 8.0$,
(d) the corresponding hysteresis ($m(t)-h(t)$) loop.

\bigskip

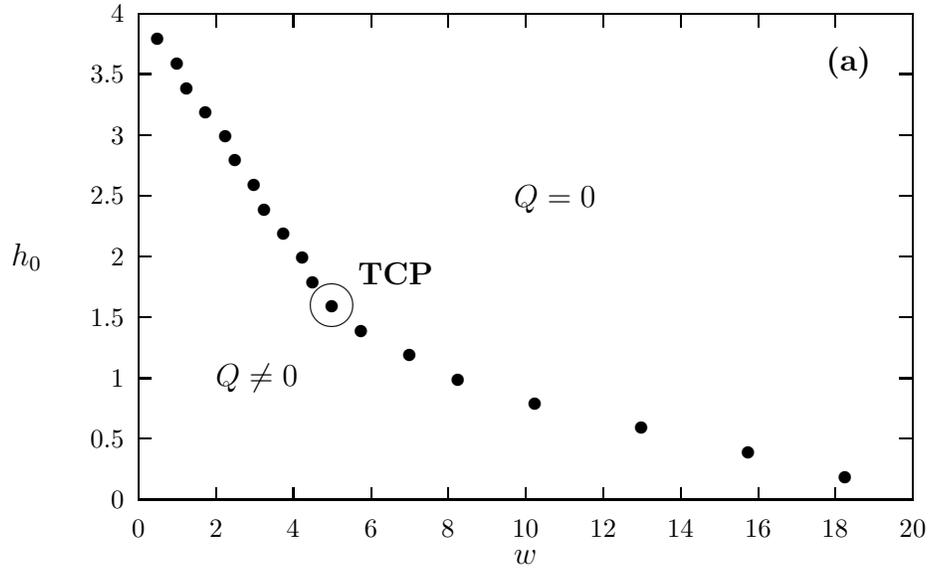
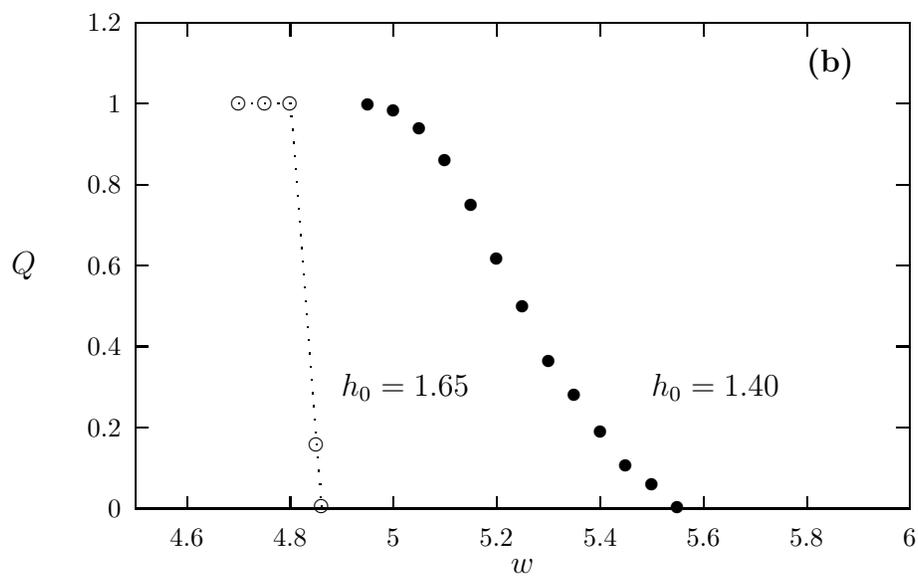
\noindent {\bf Fig. 2.} (a) The phase boundary (of the dynamic transition) 
in $w-h_0$ plane. The tricritical point (TCP) lies within the encircled
region. The boundary of the circle is the uncertainty associated in locating
the TCP, (b) two typical transitions just below and above the tricritical
point which show the different natures (discontinuous/continuous)
 of the transitions.
\end{document}